\documentclass[aps,prl,twocolumn,showpacs,superscriptaddress,letterpaper,amsmath,amssymb]{revtex4} 

\usepackage{graphicx}  
\usepackage{dcolumn}   
\usepackage{bm}        
\usepackage{amssymb}   
\usepackage{feynmf}
\usepackage{slashed}
\def\gsim{\lower0.5ex\hbox{$\:\buildrel >\over\sim\:$}}
\def\lsim{\lower0.5ex\hbox{$\:\buildrel <\over\sim\:$}}

\def \n{\noindent}

\let\n=\nu
\let\t=\tau

\newcommand{\be}{\begin{equation}}
\newcommand{\ee}{\end{equation}}
\newcommand{\bea}{\begin{eqnarray}}
\newcommand{\eea}{\end{eqnarray}}

\newcommand{\nbox}{{\,\lower0.9pt\vbox{\hrule \hbox{\vrule height 0.2 cm
\hskip 0.2 cm \vrule height 0.2 cm}\hrule}\,}}

\begin{document}

\thispagestyle{empty}
\vspace*{-3.5cm}

\vspace{0.5in}

\title{Reinterpretion of Experimental Results with Basis Templates}

\begin{center}
\begin{abstract}
Experimental analysis of data from particle collisions is typically expressed as
statistical limits on a few benchmark models of particular, often
historical, interest.  The implications of the data for other
theoretical models (current or future) may be powerful, but they cannot typically be calculated
from the published information, except in the simplest case of a
single-bin counting experiment.  We present a novel solution to
this long-standing problem by expressing the new model as a linear
combination of  models from published experimental
analysis, allowing for the trivial calculation of
limits on a nearly arbitrary model.  We present tests in simple toy experiments, demonstrate
self-consistency by using published results to reproduce other
published results on the same spectrum, and provide a reinterpretation
of a search for chiral down-type heavy quarks  ($b'$) in terms of a
search for an exotic heavy quark ($T$) with similar but distinct
phenomenology. We find $m_T>419$ GeV at 95\% CL, currently the
strongest limits if the $T$ quark decays via $T\rightarrow Wb,
T\rightarrow tZ$ and $T\rightarrow tH$.
\end{abstract}
\end{center}

\author{Kanishka Rao}
\author{Daniel Whiteson}
\affiliation{UC Irvine, Irvine, CA }

\maketitle

\section{Introduction}

Data from particle colliders may reveal  new states of matter or
evidence for new forms of interactions, or disprove theories of such
new phenomena.  When no evidence of new phenomena is seen, the
experimental collaborations who collect and analyze the data
communicate the non-observation in terms of statistical limits on
theories which predict the new states or interactions.

The space of possible theoretical models is impossibly vast;
therefore experimental results are typically communicated as limits
on a few benchmark models of particular or historical interest.
However, the
data can  provide tight constraints on many other models not included in the
experimental analysis -- such as models not yet constructed. One
solution would be for the experiments to provide a rapid mechanism for
testing new models against previously analyzed datasets.  Currently, however, the experiments are not
capable of (or perhaps not interested in) responding to an
exhaustive list of models; furthermore, experimental re-analysis
typically occurs on a timescale of weeks or months rather than hours or days.
Some ideas have been proposed to smoothen this process, such as
{\sc recast}~\cite{recast}, which elegantly connects the individual
experimental experts to the theoretical models, but does not remove
the lengthy experimental review process and so does not provide a
rapid and certain solution -- the experiment may still choose to not
provide limits on the requested models.

To date, it has been impossible for those outside the experimental
collaborations to reinterpret the published results on
benchmark models in order to set limits on new models, except in a single
restrictive case. When experimental analysis is performed as a simple selection of
events (e.g.~\cite{atlasss,atlaszz}), it can be imagined as a counting experiment and the information needed to derive limits on new models is
often included in the experimental publication. To calculate a
statistical upper limit on the number of events from a new source, $N_{\rm events}$,
which is compatible with the data, all that is needed are the expected
contributions from standard model (SM) backgrounds  (with
uncertainties) and the
observed collider event yield.  One can then calculate an upper limit
on the production cross section of the new
source, $\sigma_{\rm new}$, using

\[ N_{\rm events} = \sigma_{\rm new} \times \epsilon_{\rm selection} \times \mathcal{L} \]

\noindent
where  $\epsilon_{\rm selection}$ is the efficiency of the
experimental detection and selection and $\mathcal{L}$ is the
integrated luminosity of the dataset. Reasonable methods exist for
estimating $\epsilon_{\rm selection}$~\cite{pgs}.

Access to only single-bin analysis for reinterpretation is an unfortunately restrictive
condition, as many of the most powerful
and important results are those which perform a multi-bin fit to a
spectrum, taking advantage of distinct signal and background
shapes (e.g.~\cite{atlasbp,atlastp,cdftj}).  Use of a multi-bin histogram  can greatly
improve the sensitivity, but makes reinterpretation difficult, as one
needs access to the bin-to-bin correlations for all background
components and each of the systematic correlations~\cite{leshouches}. 
Multi-bin data may be reinterpreted directly in terms of new theoretical models
but these rely on simple detector simulation tools which may not accurately describe the reconstructed
variable. It is also difficult to describe all the systematic uncertainties, especially in a multi-bin
analysis where a systematic uncertainty may also distort the shape of the template.
Such an approach is suited for comparing features in the data against specific new models~\cite{wjjh} 
but cannot be reliably used to to calculate realistic limits in the manner of the experimental collaborations. 
The result is that a large fraction of published results are unusable for
reinterpretation.

A  solution to this problem would be experimental publication
of all the details of the limit calculation, for example the
complete likelihood used to analyze the final binned data.  This is
not currently provided by the experiments, though there are prospects
for future mechanisms~\cite{hepdata,roostats}.

We present a novel solution to this problem, which allows for the
reinterpretation of published limits derived from multi-bin analysis.  If
the binned distribution of a new theoretical model can be expressed as a linear
combination of the models for which published limits exist, a simple
relationship allows the calculation of the limit on the new model as a
simple combination of the limits on the published models.
This allows a large swath of previously un-usable experimental results
to confront a nearly arbitrary space of possible models without
involvement by the experiment.

In the following, we introduce "the basis-limit hypothesis", demonstrate
its effectiveness in simple artificial scenarios, show self-consistency by using published results to reproduce other
published results on the same spectrum, and provide a reinterpretation
of a search for chiral down-type heavy quarks  ($b'$) in terms of a
search for an exotic heavy quark ($T$) with similar but distinct phenomenology.

\begin{figure}[Ht]
\includegraphics[width=1.0\linewidth]{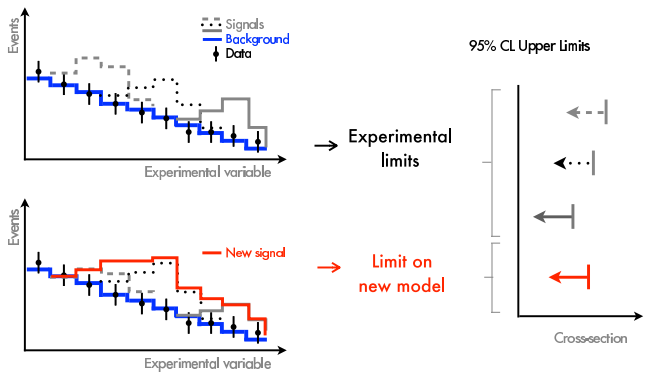}
\caption{Upper limits on the production cross section of a  new signal hypothesis can be derived if the new
  signal can be constructed as a linear combination of signals with
  existing experimental limits.}
\label{fig:diag}
\end{figure}

\section{The Basis-Limit Hypothesis}

For a specific experimental dataset and background model which is
analyzed using a binned likelihood in a variable $x$, there are a set of $n$
signal hypotheses described by binned distributions (``templates'') 
$f_1(x),f_2(x),...,f_n(x)$. In the case that the data prefer the
background model, each signal template has an associated cross-section
upper limit, which can be expressed relative to the theoretical
prediction for that signal hypothesis: $\sigma^{\rm
  limit}_i/\sigma_i^{\rm theory}$, where $\sigma_i^{\rm theory}$ is the
cross section used to normalize the template $f_i$, such that

\[  N_i^{\rm events} = \int f_i(x) = \sigma_i^{\rm theory} \times
\epsilon_i \times \mathcal{L}, \]

\noindent
and $\epsilon_i$ is the selection efficiency for the $i$-th signal
hypothesis.  

Given a new, untested signal hypothesis template, $F(x)$, which may be
expressed as a linear combination of $f_i(x)$, as

\[ F(x) = \sum_{i=0}^n a_{i} f_i(x)\ \ (a_i \ge 0), \]

\noindent
our hypothesis is that an upper limit on the cross-section of the new
signal hypothesis, $\sigma_F^{\rm limit}$ can be calculated from the
limits ($\sigma^{\rm
  limit}_i/\sigma_i^{\rm theory}$) associated with
each $f_i(x)$, as
\begin{equation}\label{eq:egl}
\frac{\sigma_{F}^{\rm limit}}{\sigma_{F}^{\rm theory}} = \left [
  \sum_{i=0}^n a_i \frac{\sigma_{i}^{\rm theory}}{\sigma_{i}^{\rm limit}}
\right ]^{-1}
\end{equation}

Stated briefly, given limits on a set of  basis templates, we claim to be able to calculate limits on any new
signal template which can be expressed as a linear combination of the basis templates. The idea is expressed diagrammatically in Fig~\ref{fig:diag}.

\section{Performance in toy examples}
\begin{figure}[Ht]
\includegraphics[width=0.5\linewidth]{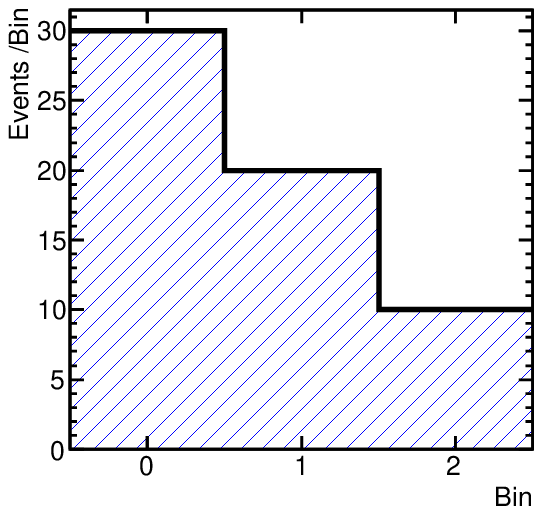}\\
\includegraphics[width=0.5\linewidth]{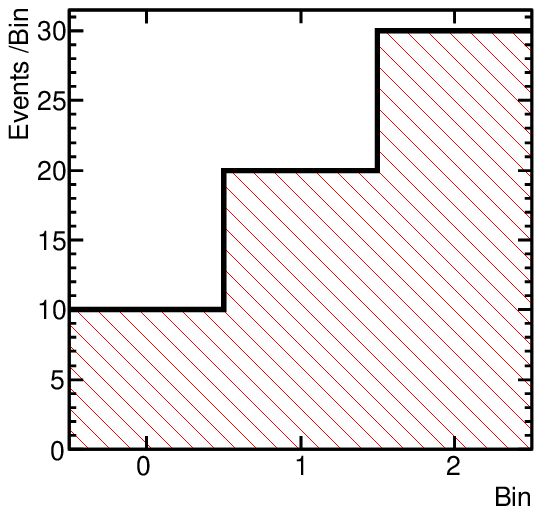}\\
\includegraphics[width=0.5\linewidth]{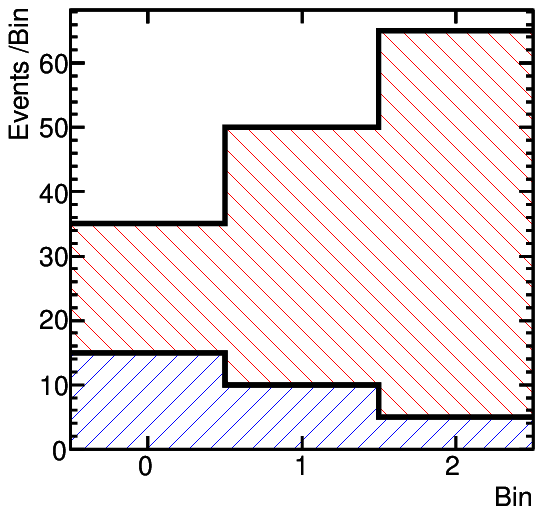}
\caption{The example toy basis templates, $f_1$ ({\it top}) ,
  $f_2$  ({\it center}) and a linear combination $F = 0.5f_1 +
  2f_2$. We derive a limit on the $F$ template using limits on
  $f_1,f_2$ and the coefficients $a_1=0.5,a_2=2.0$, see Table~\ref{tab:example}.}
\label{fig:f123}
\end{figure}

We demonstrate the application of the basis-limit hypothesis using a
simple toy scenario: a 3-bin analysis with two signal hypotheses,
shown in Figure~\ref{fig:f123}.  The two signal templates are the
basis templates and are generated with an arbitrary but equal signal
cross-section, $\sigma^{\rm theory}$.

We use the CLs technique~\cite{cls1,cls2} to estimate cross-section
upper limits on each of the basis templates relative to a flat background.  For a new signal hypothesis,
$F$, expressed as a linear combination of the templates, we
predict the limit on $F$ using the limits on the basis templates, as
shown in Equation~\ref{eq:egl}.  Table~\ref{tab:example} gives an example
for a single case.

\begin{table}
\caption{Demonstration of basis-limit application in a toy scenario. A
  new signal hypothesis, $F$, is formed from a linear combination of
  $f_1$ and $f_2$  using the coefficients $a_i$ (see Fig.~\ref{fig:f123}). The limits
  ($\sigma^{\rm limit}_i/\sigma_i^{\rm theory}$) on the
  $f_i$ and the coefficients $a_i$ can be used to predict the limit on
  the new hypothesis, $F$, using Equation~\ref{eq:egl}. The predicted
  limit is confirmed by an explicit calculation of the limit using the
  $F$ template.}
\label{tab:example}
\begin{tabular}{lrrr}
\hline \hline
 & $f_1$ & $f_2$  & $F$ \\ \hline
$a_1$ & 1.0 & -- & 0.5 \\
$a_2$ & -- &  1.0 & 2.0 \\ \hline
$\sigma^{\rm limit}_i/\sigma_i^{\rm theory}$ (meas) & 0.27 & 0.27 &
{\bf 0.11} \\
$\sigma^{\rm limit}_i/\sigma_i^{\rm theory}$ (pred) & -- & -- &
{\bf 0.11} \\
\hline \hline
\end{tabular}
\end{table}

\begin{figure}[Ht]
\includegraphics[width=0.98\linewidth]{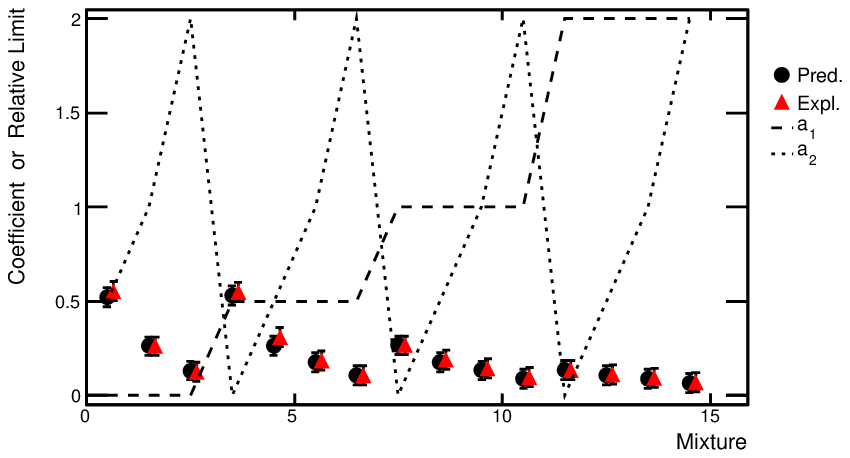}
\includegraphics[width=0.98\linewidth]{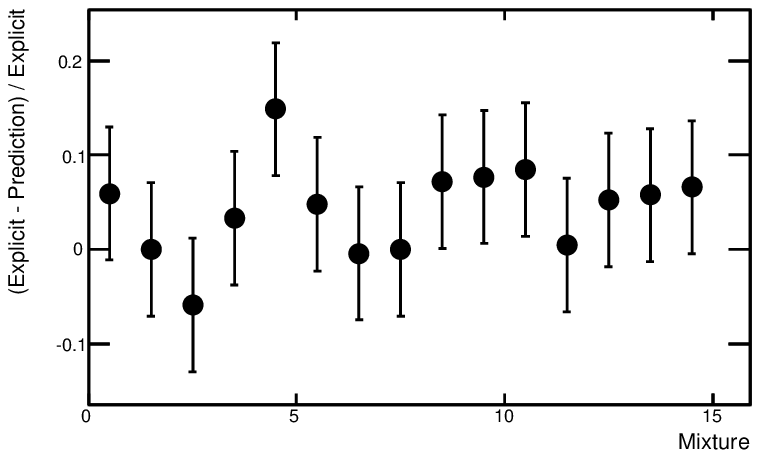}
\caption{ Tests of the basis-limit hypothesis in a toy scenario
  involving two signal templates, see Figure~\ref{fig:f123}. For a
  new shape $F = a_1 f_1 + a_2f_2$ and a variety of $a_1,a_2$
  mixtures, we compare the relative upper limit on the model described
  by the $F$ template as calculated explicitly
  to that predicted by the basis-limit formula, which uses only the
  limits on $f_1,f_2$ and the coefficients $a_1,a_2$.  Top pane shows
  the explicit and predicted upper limits for varying $a_1,a_2$; bottom pane
  shows the difference in the explicit and predicted upper limits. Statistical
  uncertainties come from the pseudo-experiments required in the limit
calculation.}
\label{fig:F}
\end{figure}

A comprehensive test, scanning many possible values of $a_i$, shows
that the basis-limit predictions are robust in this toy example, see Fig.~\ref{fig:F}.

\section{Possible shortcomings}

\subsection{Limited basis templates}

The basis-limit hypothesis is not universally robust: it cannot be
blindly applied to every published result or
 completely arbitrary new models. The primary limitation is whether
enough example shapes are presented to form a basis to describe a new test signal.

 As an extreme example to illustrate the point, consider a two-bin analysis
with large background rates in each bin: $f_{bg}(x) = (10^3,10^3)$, but
large uncertainty: $\Delta f_{bg}(x) = (10^2,10^2)$.
 A two-bin signal template with signal
isolated in just one bin, $f_1(x) = (100,0)$, may have reasonable sensitivity, as the
background uncertainty can be reduced by a fit to the observed data in the signal-depleted
bin. But, if we were to make a poor choice of basis templates, in which each bin
kept significant signal contributions: $f_1(x) = ( 101,99),f_2(x) =
(99,101) $, no reduction in the background uncertainty would be possible for
either signal template, leading to weak limits for both.  There is no
positive set of coefficients $a_i$ which can be combined to describe a
signal hypothesis with signal isolated in one bin; these templates
fail to capture the real power of the data set.

The basis-limit hypothesis implicitly assumes that the constraints on
the background systematics obtained when determining limits for each
of the basis templates is comparable to the constraints that would
be obtained when determining limits using the new signal.  This
assumption is valid if, as is often the case, the background
systematics are predominantly constrained in a background-rich region
where both the basis template and new-signal are always small.

Another extreme example where the basis-limit hypothesis would fail is
the case of a new signal template which is identical to a published
template, but with larger or different systematic uncertainties.

\subsection{Approximate Efficiency Calculations}

Any reinterpretation of published data in terms of a new theoretical
model requires an estimate of the efficiency, $\epsilon_{selection}$
of the experiment to detect and select events from the proposed new
source. While the most accurate estimate of $\epsilon_{selection}$ can
only be performed by the experimental collaboration, often via use of
their private official {\sc geant}-based~\cite{geant} detector simulation programs, there are well-established public
tools, such as {\sc pgs}~\cite{pgs}, which provide estimates via a
parametric simulation with reasonable accuracy ($5-20\%$ relative) in
most regimes.

One application of the basis-limit approach is to build templates of
the new theory using the available public simulation programs and express
them as linear combinations of the published templates produced by the
experiments with their private simulation programs.  This incurs the
same acceptable level of uncertainty in each bin as in the well-established
single-bin case.  The approximation of $\epsilon$ results in an
approximate determination of the $a_i$ coefficients.

 To calculate the limits on a new theory, the $\epsilon$ values are
 not directly needed -- all that is required
are the coefficients $a_i$ and the limits on each template $f_i$.  If
the new theory templates and the basis templates are built using the
same approximate public simulation, many of the approximations may
cancel. For example, if $\epsilon^{F}_{\rm approx}/\epsilon^F_{\rm official} =
  \epsilon^{f_i}_{\rm approx}/\epsilon^{f_i}_{\rm official}$ then it may be possible to calculate the $a_i$ without incurring
approximations due to $\epsilon$.  In addition, it
is possible with this approach to make templates in cases when the published analysis
has limits  quoted for many cases but only a few example templates
(e.g.~\cite{atlaszp}).

\section{ Tests with published limits }
The toy scenario above demonstrates the validity of the basis-limit
hypothesis  when the new signal is exactly a
linear combination of previously examined signals.  In this section,
we demonstrate the use of basis limits in realistic scenarios using published experimental results.

\subsection{ Same-sign dileptons at CDF }

A critical test of the basis-limit hypothesis is a comparison of limits
predicted using Equation~\ref{eq:egl} with limits derived by the
experiment itself.  This requires a pair of experimental limits which
use identical datasets, selections and background models.  One such pair
is a search at CDF in same-sign dileptons with jets and missing
energy in 6.1 fb$^{-1}$; the dataset was used to extract limits on
supersymmetry~\cite{cdfsssusy} and same-sign top-quark pair
production~\cite{cdfsstops}. Both analyses use $H_{T}$, the scalar sum of transverse momenta
  of jets and leptons, as the
discriminating variable.

\begin{figure}[Ht]
\includegraphics[width=0.5\linewidth]{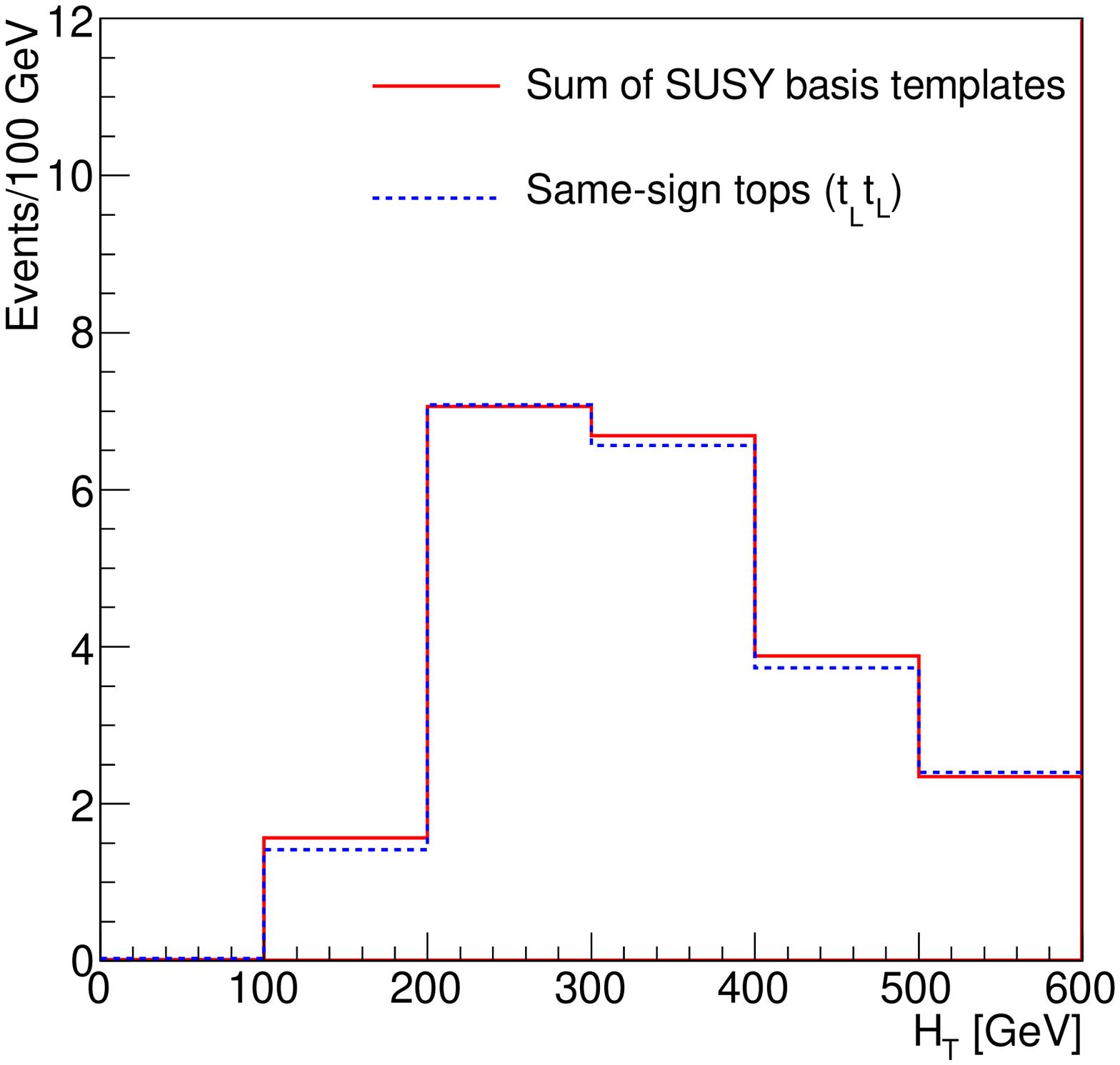}\\
\includegraphics[width=0.5\linewidth]{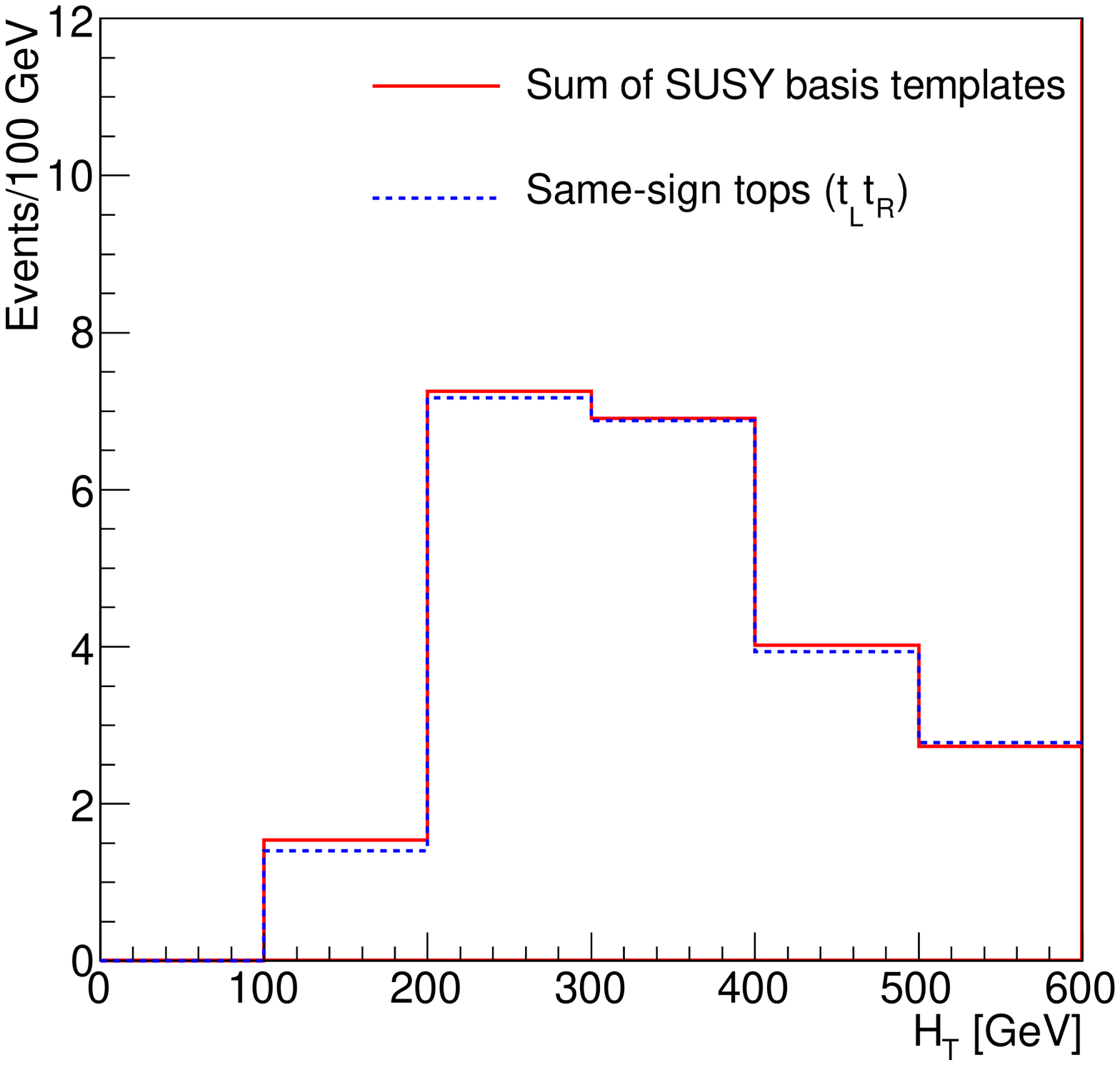}\\
\includegraphics[width=0.5\linewidth]{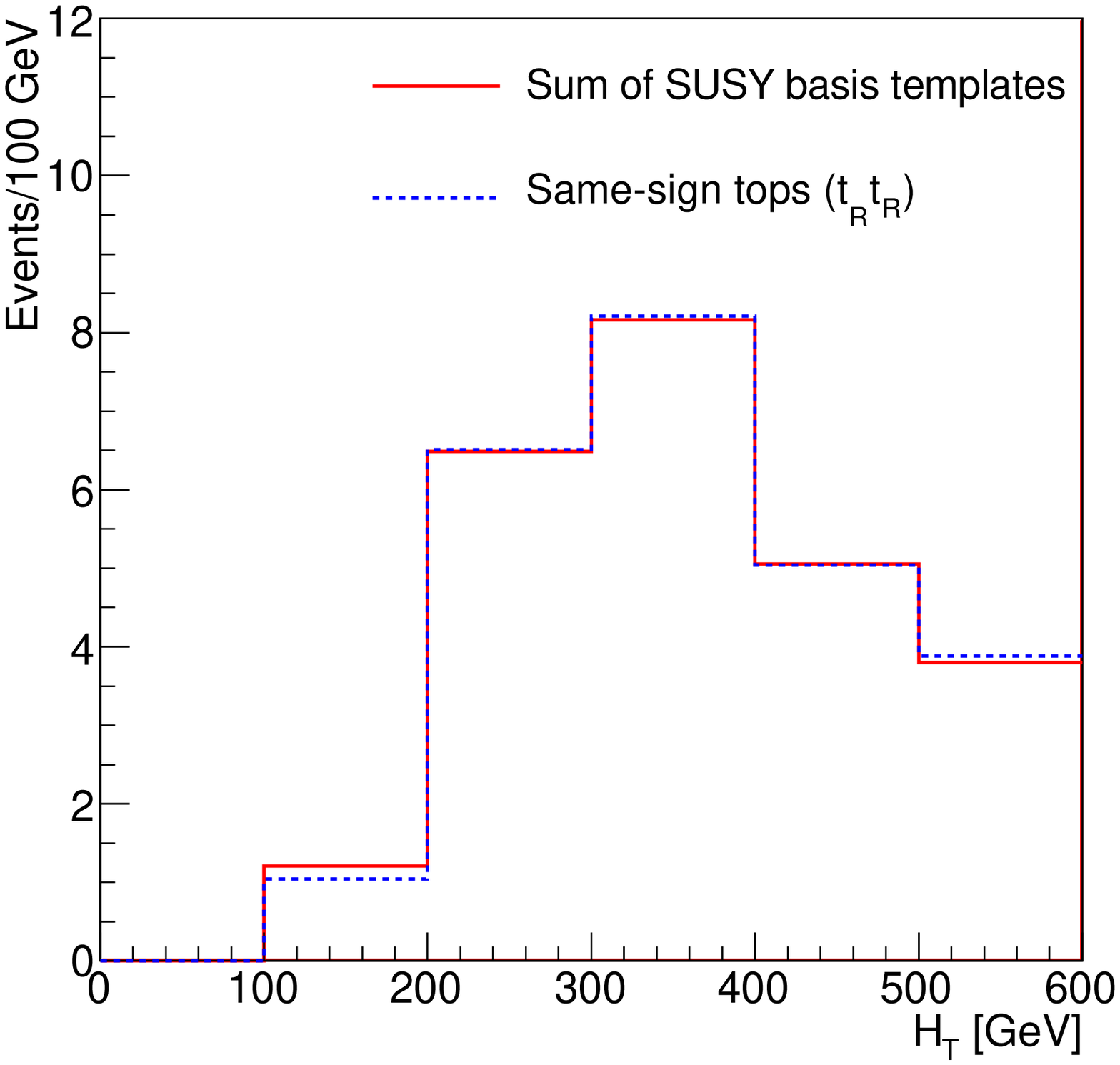}
\caption{ Distribution in event $H_T$ (the scalar sum of transverse momenta
  of jets and leptons) from same-sign top-quark pair production (blue, dashed)
  expressed as a linear combination of distributions from gluino-pair
  production and squark-pair production. Panes show
  show the $t_Lt_L$ (top), $t_Lt_R$ (center) and $t_Rt_R$ (bottom) chirality configurations. }
\label{fig:susy}
\end{figure}

We form linear combinations of the SUSY templates to reproduce the
same-sign top-quark templates, see Figure~\ref{fig:susy}. The
coefficients $a_i$ and the published limits on each of the SUSY
templates can then be used to predict the limits
on same-sign top-quark production, see Table ~\ref{tab:susy}. The
predicted limits agree with the published results in each case.

\begin{table} 
\caption{ Prediction of limits on same-sign top quark pairs in various
chirality configurations derived from limits on supersymmetric
production of gluinos or squarks. Coeffients of the supersymmetric
particle templates are denoted by the particles and mass heirarchy,
all in units of GeV.
Note that coefficients are individually quoted only for the three most
significant basis templates; the sum of the 30
remaining coefficients, $a_{\rm other}$ is also shown.}
\label{tab:susy}
\begin{tabular}{ l c c c}
\hline\hline
       & $t_Lt_L$ & $t_Lt_R$ & $t_Rt_R$ \\ \hline
$a_{\tilde{g}(m=200),\chi^+(m=75),\chi^0(m=50)}$ & 0.061 & 0.074& 0.098\\
$a_{\tilde{g}(m=300),\chi^+(m=150),\chi^0(m=100)}$ & 0 & 0 & 0.015\\
$a_{\tilde{q}(m=200),\chi^+(m=75),\chi^0(m=50)}$ & 0.040& 0.035& 0.011\\
$\Sigma a_{\rm other}$ & 0.100  & 0.068 & 0.700\\
\hline
Experiment Results~\cite{cdfsstops} [fb] & 54 & 51 & 44 \\
Our Prediction [fb] & 53.7 & 50.9 & 44.1 \\
\hline\hline
\end{tabular}
\end{table}

\subsection{ Self-consistency test}

Pairs of published experimental results with identical selection,
dataset and backgrounds but distinct signal hypotheses are quite
rare.  However, we can probe the basis-limit performance in realistic
scenarious using a self-consistency test. 

In a set of $N$ signal templates, we can
attempt to describe the $i$-th template using the other $N-1$
templates. Given the published limits on the $N-1$ templates, we can
predict the limit on the $i$-th template and compare it to the
published limit.

\begin{figure}[Ht]
\includegraphics[width=0.6\linewidth]{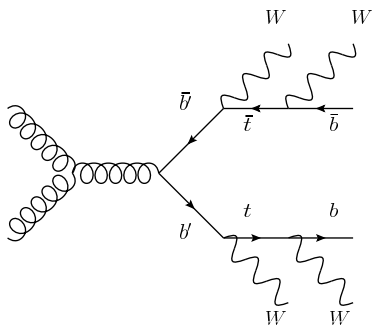}
\caption{Pair-production of heavy down-type $b'$ quarks with decay
  $b'\rightarrow Wt$.}
\label{fig:bp_diag}
\end{figure}

\begin{figure}[Ht]
\vspace{3cm}
\includegraphics[width=0.48\linewidth]{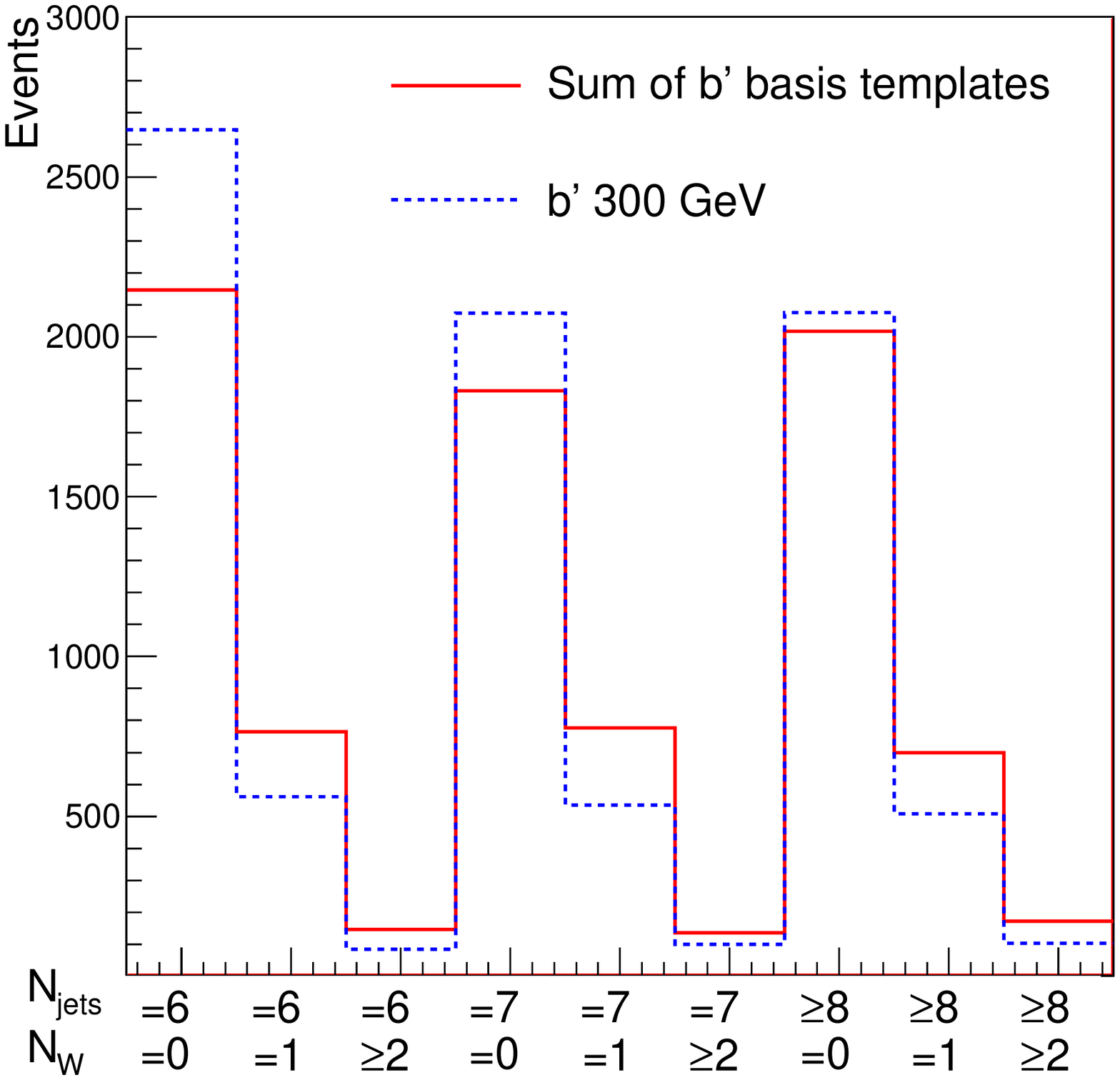}
\includegraphics[width=0.48\linewidth]{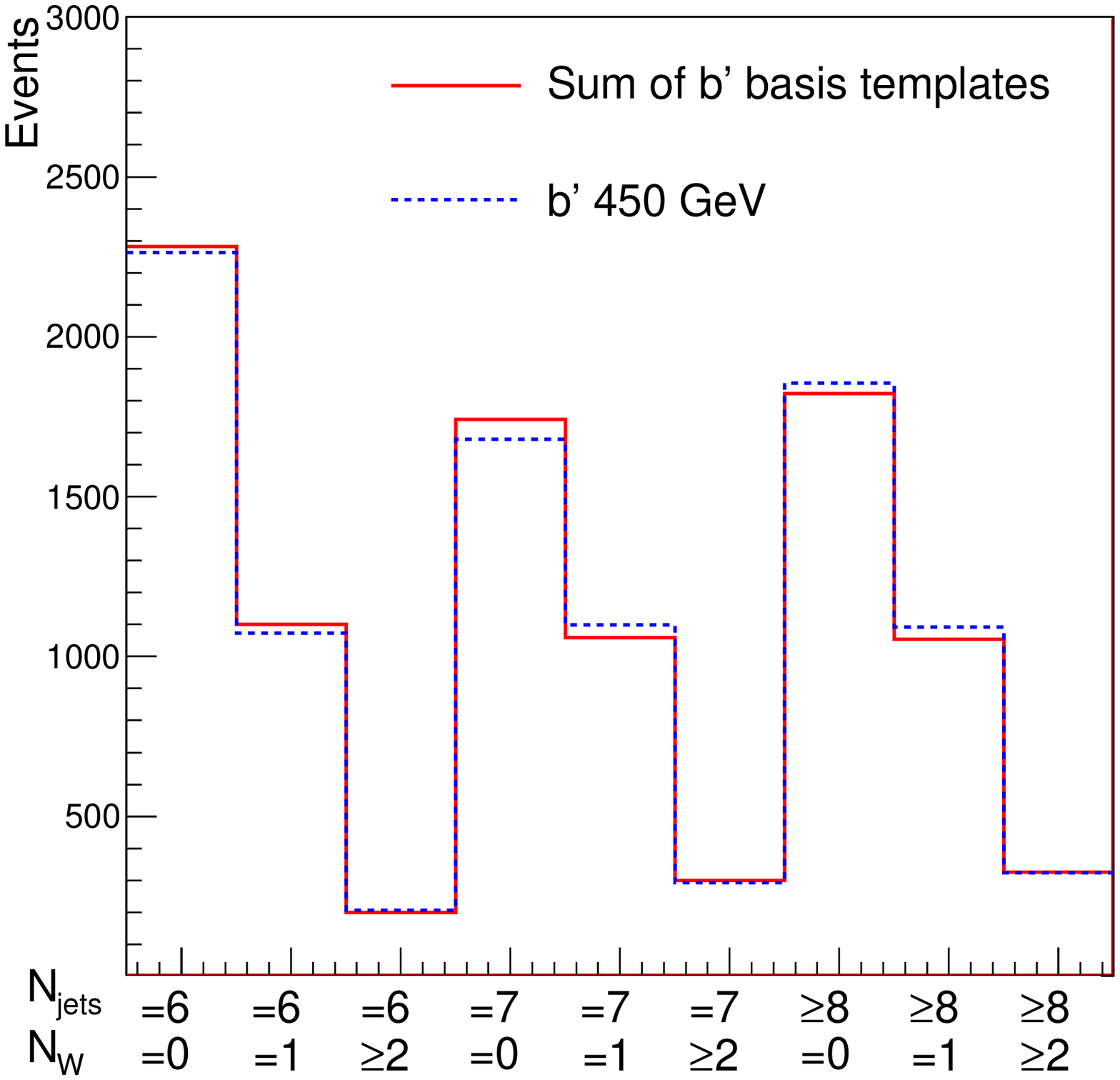}
\caption{ Jet multiplicity and hadronically-decaying $W$ multiplicity
  for $b\rightarrow tW$ pair production  GeV (blue, dashed)
  expressed as a linear combination (red, solid) of distributions from other $b'$
  masses, see Table~\ref{tab:bprime}. Left is for $m_{b'}= 300$ GeV; right
  is for $m_{b'}= 450$ GeV. For the edge case of $m_{b'}= 300$ GeV, the sum of basis templates
  is not exact, due to the lack of templates at lower masses.  }
\label{fig:njnw}
\end{figure}

The ATLAS collaboration reported a search for heavy fourth-generation
down-type chiral quarks ($b'$) using 1 fb$^{-1}$~\cite{bprime}.   The
$b'$ decays via $tW$, leading to a final state with four $W$ bosons
and two $b$ quarks, see Fig.~\ref{fig:bp_diag}. The ATLAS search makes use
of a novel technique for tagging boosted $W$ bosons by searching for
jet pairs with small angular separation. The analysis variable is the
jet multiplicity and $W$ boson multiplicity, see Figure~\ref{fig:njnw}.

We generate $b'\rightarrow tW$ using {\sc madgraph}~\cite{madgraph},
use {\sc pythia}~\cite{pythia} to model showering and hadronization,
and {\sc pgs} to describe the detector response.  Details of the construction of the templates and the resulting limits
are given in Table~\ref{tab:bprime}. 

\begin{figure}[Ht]
\includegraphics[width=0.65\linewidth]{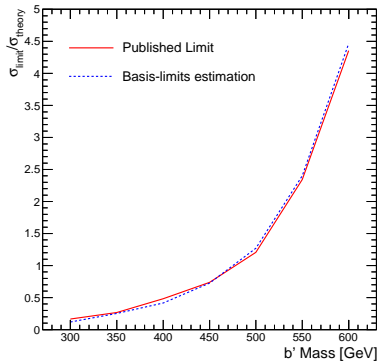}\\
\caption{ Prediction of limits on $b'\rightarrow tW$ decays at
  ATLAS. Limits at each mass are predicted by expressing the signal
  model at that mass in terms of signal models at other masses; these
  predicted limits agree well with the limits reported by
  ATLAS~\cite{atlasbp}, see Table~\ref{tab:bprime}.}
\label{fig:bp_self}
\end{figure}

Figure~\ref{fig:bp_self} shows that the basis-limit estimation is
reliable for this application.  At the lower boundary, $m_{b'} = 300$
GeV, it is difficult to find coefficients which give an accurate
description, see Figure~\ref{fig:njnw}.

\begin{table}
\caption{ Details of the self-consistency test using $b'\rightarrow
  tW$ decays at ATLAS.  The template at each specific  mass  is expressed as a linear
  sum of templates at other masses; the predicted limit is then
  compared to the explicit calculation.}
\label{tab:bprime}
\begin{tabular}{lrrrrrrr}
\hline\hline
& $b'_{300}$ & $b'_{350}$ & $b'_{400}$ & $b'_{450}$ & $b'_{500}$ & $b'_{550}$ & $b'_{600}$ \\ \hline
$a_{b' 300}$  & -  & 0.19  & 2e-3  & 0  & 0  & 0  & 0  \\
 $a_{b' 350}$  & 2.29  & -  & 0.16  & 0  & 0  & 0  & 0  \\
 $a_{b' 400}$  & 0  & 1.10  & -  & 0.23  & 0.03  & 0  & 0  \\
 $a_{b' 450}$  & 0  & 0.32  & 1.31  & -  & 0.09  & 0.03  & 0  \\
 $a_{b' 500}$  & 0  & 0.01  & 0  & 0.35  & -  & 0.08  & 0.02  \\
 $a_{b' 550}$  & 0  & 0.01  & 2e-3  & 1.10  & 0.40  & -  & 0.48  \\
 $a_{b' 600}$  & 0  & 1e-3  & 0  & 0.59  & 1.92  & 1.35  & -  \\\hline
$\sigma^{\rm limit}_i/\sigma_i^{\rm theory}$ (meas) & 0.16 & 0.27 & 0.48 & 0.74 & 1.2 & 2.3 & 4.3\\
$\sigma^{\rm limit}_i/\sigma_i^{\rm theory}$ (pred) & 0.12 & 0.26 & 0.42 & 0.73 & 1.3 & 2.4 & 4.5 \\
\hline\hline
\end{tabular}
\end{table}

\section{New limits on a  heavy exotic quark: $T$}

Having shown the validity of the basis-limit hypothesis, we provide a
demonstration of the calculation of limits on an untested signal
hypothesis using published experimental results.

\begin{figure}[Ht]
\includegraphics[width=0.9\linewidth]{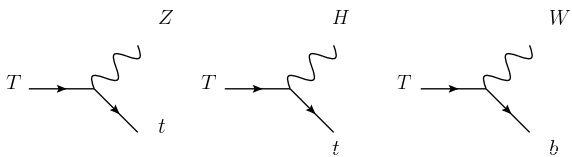}
\caption{Decay modes of a heavy exotic quark, $T$}
\label{fig:tp_diag}
\end{figure}

An exotic heavy quark $T$ may decay as \mbox{$T\rightarrow Wb$},
\mbox{$T\rightarrow tZ$}, or \mbox{$T\rightarrow tH$}, see Fig.~\ref{fig:tp_diag}.  The signature
of the decay is similar to that of the $b'$ model, involving hadronic
decays of boosted bosons ($W,Z,H$) and top quarks~\cite{tprime}.  The
CMS collaboration analyzed data with 1.14 fb$^{-1}$ of integrated
luminosity and excluded at 95\% CL such a $T$ quark below $m_T = 475$ GeV
assuming BR($T\rightarrow tZ=100$\%)~\cite{cmstp}.  In the more likely
configuration with other decay modes available and BR($T\rightarrow tZ
\leq 25$\%) (see ~\cite{tprime}), the CMS limit would be
significantly weaker, perhaps $m_T > 250$ GeV.

\begin{figure}[Ht]
\vspace{3cm}
\includegraphics[width=0.65\linewidth]{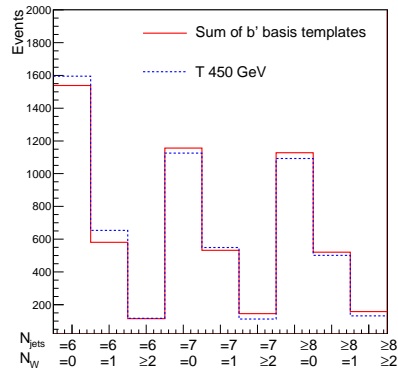}
\caption{Jet multiplicity and hadronically-decaying $W$ multiplicity
  for exotic heavy quark $T$  pair production and decay with $m_{T}=450$ GeV (blue, dashed)
  expressed as a linear combination of distributions from
  $b'\rightarrow tW$ model with varying masses, see
  Table~\ref{tab:tprime}. We include all $T$ decay modes, see Fig.~\ref{fig:tp_diag}.}
\label{fig:tp_njnw}
\end{figure}

We combine all decay
modes together to maximize the expected yield and to be sensitive to the broader model.
As before, we construct templates for $T$ as linear combinations of the
existing $b'$ templates (Fig.~\ref{fig:tp_njnw}) and use Equation~\ref{eq:egl} to calculate new
limits on the pair-production of $T$ at the
LHC. 

\begin{figure}[!Ht]
\includegraphics[width=0.63\linewidth]{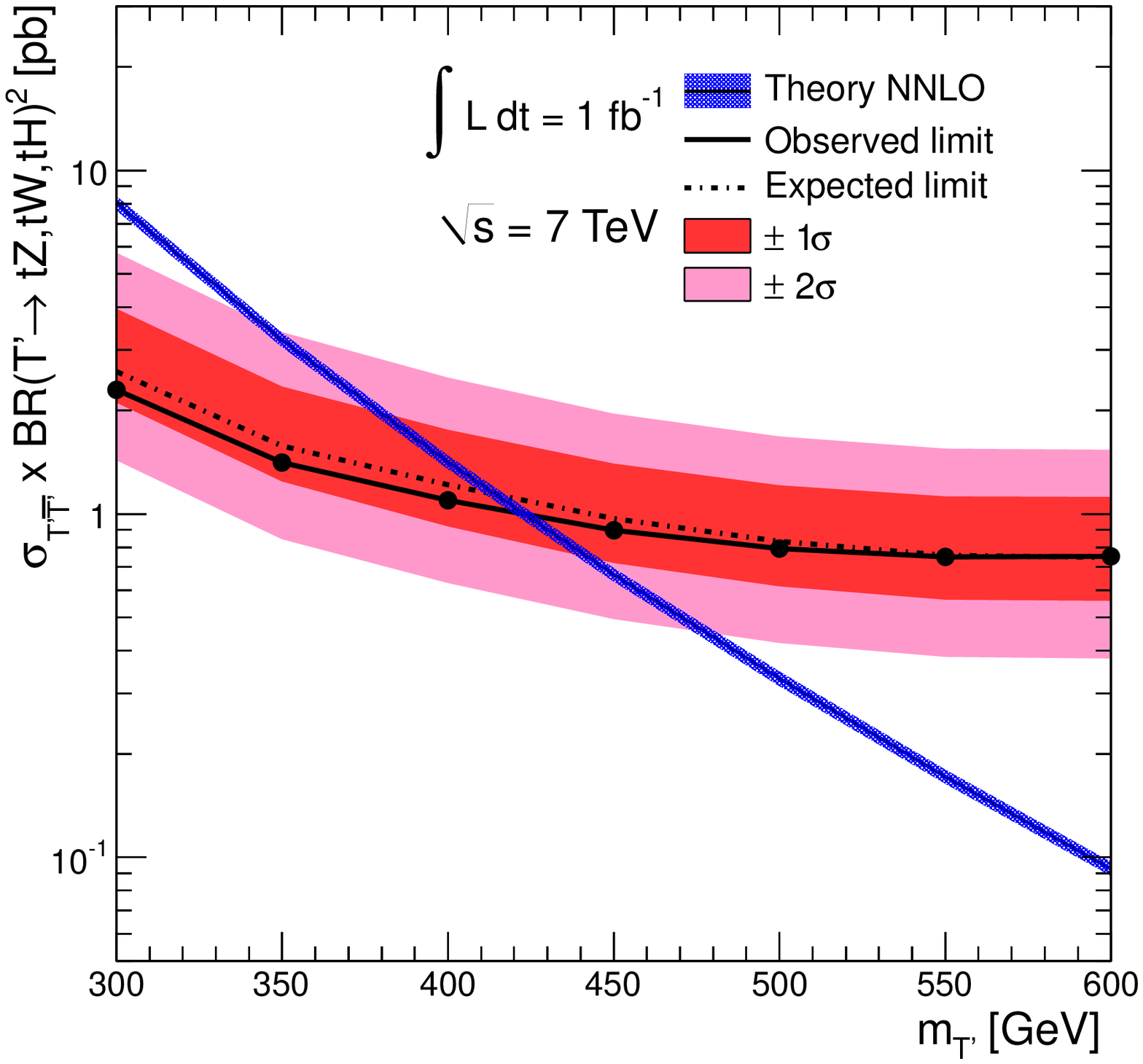}\\
\includegraphics[width=0.63\linewidth]{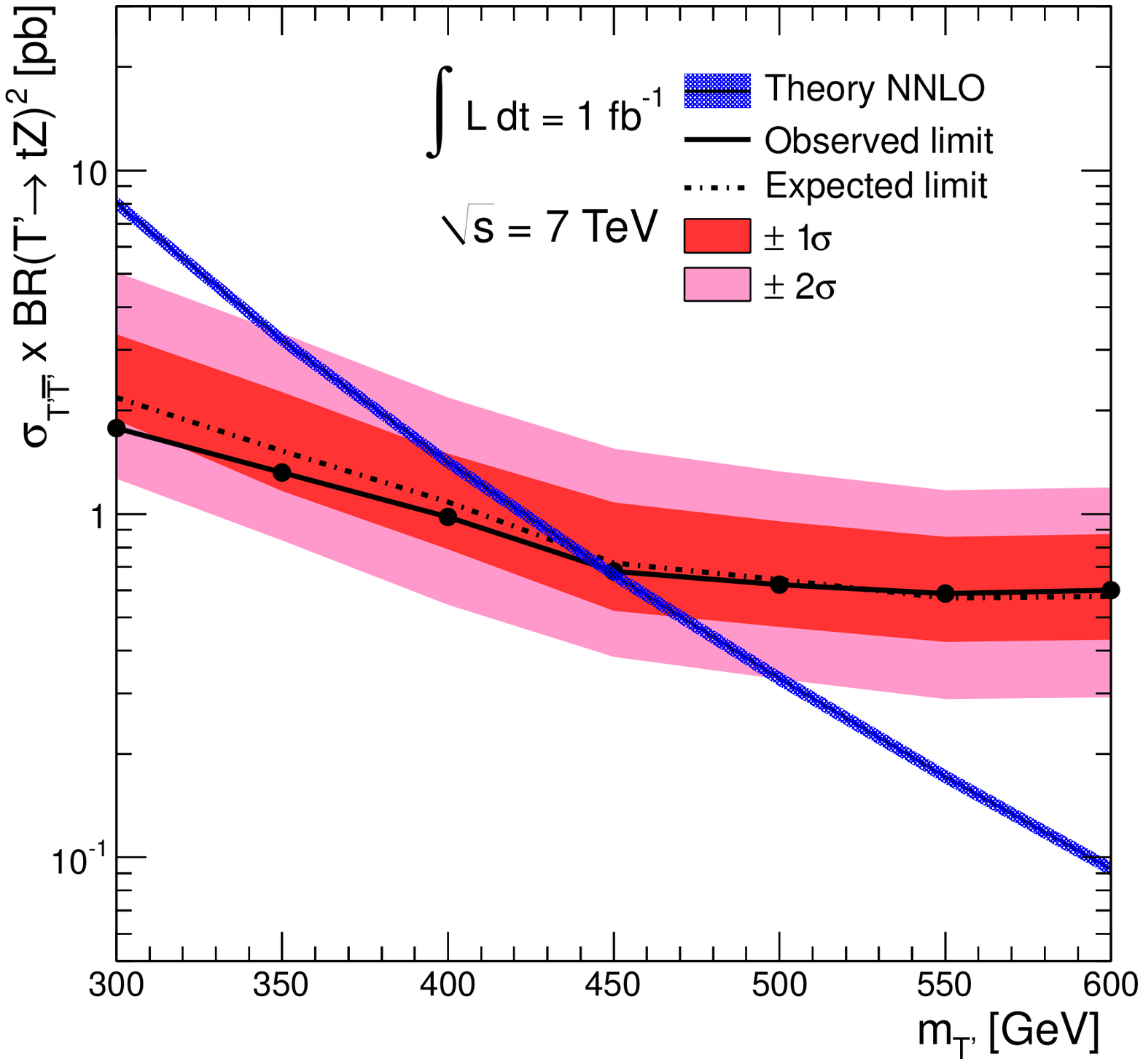}
\caption{Upper limits at 95\% CL on the production of an exotic heavy
  quark $T$. Top includes all decays (see Fig.~\ref{fig:tp_diag});
  bottom includes only $T\rightarrow tZ$ decays. The theoretical
  prediction is at approximate next-to-next-to-leading-order~\cite{d4xs}.}
\label{fig:tp_limits}
\end{figure}

Table~\ref{tab:tprime} shows the coefficients and calculated
limits, also shown in Fig.~\ref{fig:tp_limits}. Using an approximate
next-to-next-to-leading-order calculation~\cite{d4xs} of the $T$
production cross-section, our cross-section upper limit excludes a $T$ quark
with mass $m_T>419$ GeV, despite the low branching ratio,
BR($T\rightarrow tZ$)=15\% at this $m_T$.

In addition, we repeat the study using only $T\rightarrow tZ$ decay
modes, as these are most similar to the $b'\rightarrow tW$ mode
originally analyzed by ATLAS, see Fig~\ref{fig:tp_limits}.

\begin{table}
\caption{ Details of the predicted limit on $T$ pair-production and
  decay,  using basis templates from $b'\rightarrow
  tW$ decays at ATLAS.  We include all $T$ decay modes, see Fig.~\ref{fig:tp_diag}.}
\label{tab:tprime}
\begin{tabular}{lrrrrrrr}
\hline\hline
&$T_{300}$ & $T_{350}$ & $T_{400}$ & $T_{450}$ & $T_{500}$ & $T_{550}$ & $T_{600}$ \\ \hline
 $a_{b' 300}$  & 0.39  & 0.20  & 0.07  & 0.02 & 0.01  & 2e-3  & 1e-3  \\
 $a_{b' 350}$  & 0  & 0  & 0  & 0  & 0  & 0  & 0  \\
 $a_{b' 400}$  & 0  & 0  & 0  & 0  & 0  & 0  & 0  \\
 $a_{b' 450}$  & 0  & 5e-3  & 4e-3  & 3e-3  & 1e-3  & 0  & 0  \\
 $a_{b' 500}$  & 0.01  & 3e-3  & 1e-3  & 3e-3  & 0  & 0  & 0  \\
 $a_{b' 550}$  & 0.24  & 0.94  & 1.24  & 0.78  & 0.39  & 0.13  & 0.04  \\
 $a_{b' 600}$  & 4.35  & 2.88  & 1.46 & 1.11  & 0.87  & 0.68  & 0.42  \\\hline
$\sigma^{\rm limit}_i/\sigma_i^{\rm theory}$ (pred) &0.29 & 0.44 & 0.77  & 1.35 & 2.39  & 4.34 & 8.10\\
$\sigma^{\rm limit}$ [pb] (pred) & 2.29 &  1.40 & 1.09  & 0.89  & 0.79  & 0.75  & 0.75  \\
\hline\hline
\end{tabular}
\end{table}

\section{Limitations and Generalizations}

While the basis-limit hypothesis is an intuitive and effective
construction, it is a heuristic formula. We do not provide its
derivation from basic statistical axioms.  

In some scenarios, it may fail to provide an accurate prediction of
the limit on a new model, as mentioned above.  Indeed, we have
observed some artificial scenarios in which the basis templates
have very little overlap that a bias may occur in the prediction.  Some
corrections may be calculable in this scenario, leading to
modification of the $a_i$ coefficients based on the {\it a priori}
overlap of the basis templates. In most cases, where the signal templates come from new physics
processes with single slow-varying parameter, such as the mass of a new particle, the templates have substantial overlap and the correction is negligible.

Alternatively, we might express Eq.~\ref{eq:egl} in another form, as 
\[
\frac{\sigma_{F}^{\rm limit}}{\sigma_{F}^{\rm theory}} = \left [
  \sum_{i=0}^n x_i \kappa_{i} 
\right ]^{-1}
\]
where $x_i$ is the bin content of the $i$-th bin for an $n$-bin analysis, and $\kappa_i$ is an
unknown constant which depends only on the background and data in the
$i$-th bin.  Rather than expressing the new signal in terms of
basis templates, we could solve for the $\kappa_i$ given a set on $n$
limits on $n$ signal templates.  This would allow the calculation of a
limit on an arbitrary signal template without concern for an overlap
correction as discussed above. We leave this for future investigation.

\section{Conclusions}

The basis-limit hypothesis provides a tool for reinterpretting the
results of experimental analysis using multi-bin data. Previously,
only single-bin analyses could be reinterpreted.  

Some technical hurdles remain; for example, if the published analysis uses a
complex technique (such as a multi-variate analysis tool) and does not
publish enough detail, then the selection cannot be reproduced. This
also applies to a single-bin analysis.

Superior solutions to the one we propose here are:
\begin{itemize}
\item Archiving and streamlining by the experiments of published
  analysis, allowing for a rapid re-interpretation in terms of a new
  model. This has the disadvantage that it places the burden on the
  experiments.
\item Publication by the experiments of all of the details necessary
  to reproduce the analysis. This has the disadvantage that it
  requires use of an approximate publicly-available simulation.
\end{itemize}

As neither of these are currently available, the basis-limit approach  makes a wide
range of results available for constraining current and future models.

We use this approach to interpret an ATLAS search for $b'\rightarrow Wt$
to set the strongest limit on an exotic heavy quark $T$ which decays
$T\rightarrow tZ,th,Wb$ at $m_T>419$ GeV at 95\% confidence level.

\section{Acknowledgements}

We thank Michael Mulhearn, Jeffrey Streets, Kathy Copic, Kyle Cranmer, Nadine Amsel, Matthew Relich and Eric Albin for useful conversations, and Graham
Kribs and Adam Martin for the $T$ quark model and technical support.
The authors are supported by grants from the Department of Energy
Office of Science and by the Alfred P. Sloan Foundation.


\begin{thebibliography}{99}

\bibitem{recast}   K.~Cranmer and I.~Yavin,
  JHEP {\bf 1104}, 038 (2011)
  [arXiv:1010.2506 [hep-ex]].


\bibitem{atlasss} ATLAS Collaboration, arXiv:1202.5520 (2012).

\bibitem{atlaszz} ATLAS Collaboration, arXiv:1203.0718 (2012).

\bibitem{pgs} J. Conway,  {\tiny \texttt{http://www.physics.ucdavis.edu/\~conway/research/software/pgs/pgs.html}}.

\bibitem{atlasbp} ATLAS Collaboration, arXiv:1202.6540 (2012).

\bibitem{atlastp} ATLAS Collaboration, arXiv:1202.3389 (2012).

\bibitem{cdftj} CDF Collaboration, arXiv:1203.3894 (2012).

\bibitem{leshouches} S. Kraml {\it et al.}, arxiv:1203.2489 (2012).

\bibitem{wjjh} Q. H. Cao {\it et al.}, J. High Energy Phys. 08 (2011) 002.

\bibitem{hepdata} The HepData Project, \texttt{http://durpdg.dur.ac.uk/}.

\bibitem{roostats} L. Moneta {\it et al.}, arXiv:1009.1003 (2010).

\bibitem{cls1} {A. Read},   J. Phys. G: Nucl. Part. Phys. {\bf 28}, 2693 (2002);
\bibitem{cls2} {T. Junk},  Nucl. Instrum. Methods A {\bf 434}, 425
  (1999).

\bibitem{geant} S. Agostinelli {\it et al.}, Nucl. Inst.  Meth. {\bf A}506 (2003) 250-303.

\bibitem{atlaszp} The ATLAS Collaboration, Phys.Rev.Lett. 107 (2011) 272002.

\bibitem{cdfsssusy} CDF Collaboration, CDF 10465 (2011).
\bibitem{cdfsstops} CDF Collaboration, CDF 10466 (2011).
\bibitem{bprime} ATLAS Collaboration, arXiv:1202.6540 (2012).


\bibitem{madgraph}
  J.~Alwall, M.~Herquet, F.~Maltoni, O.~Mattelaer and T.~Stelzer,
  JHEP {\bf 1106}, 128 (2011)
  [arXiv:1106.0522 [hep-ph]].

\bibitem{pythia}
  T.~Sjostrand, S.~Mrenna and P.~Z.~Skands,
  JHEP {\bf 0605}, 026 (2006)
  [hep-ph/0603175].


\bibitem{tprime}   G.~D.~Kribs, A.~Martin and T.~S.~Roy,
  Phys.\ Rev.\ D {\bf 84}, 095024 (2011)
  [arXiv:1012.2866 [hep-ph]].

\bibitem{d4xs} M. Aliev {\it et al.}, Comput. Phys. Commun. {\bf 182}, 1034 (2011).

\bibitem{cmstp} CMS Collaboration, Phys. Rev. Lett. 107, 271802 (2011).

\end{thebibliography}
\end{document}